# DIRECT CHAOTIC COMMUNICATIONS IN MICROWAVE BAND

A. S. Dmitriev, A. I. Panas, S. O. Starkov

*Abstract*--We discuss the concept of Direct Chaotic Communication (DCC). The scheme is based on the following ideas: (1) the chaotic source generates chaotic oscillations directly in the specified microwave band; (2) information component is input by means of formation of the appropriate stream of chaotic radio pulses; (3) envelope detection is used. The principle of communication scheme is confirmed experimentally in microwave band. The transmission rates of 3.34, 10.0 and up to 70.0 Mbps are demonstrated.
his only looks like the abstract.

## I. INTRODUCTION

Investigations of the last years [1]–[8] indicate that dynamic chaos has a number of features that make it attractive for use in communication systems as carrier or modulated oscillations. Among these features, there are: generation of complicated oscillations in simple-structure devices; many different chaotic modes may be realized in a single source; chaotic modes can be controlled by small variations of the system parameters; variety of methods for putting information in the chaotic signal; self-synchronization of transmitter and receiver; confidentiality of communications.

Indeed, with contemporary communication tools one can and successfully solves most of the occurring problems using conventional methods, based on regular signals or, if spectrum spreading is necessary, on pseudo-random signals. However, what is attractive in the use of dynamic chaos is that the set of requirements to contemporary and promising communication systems can be satisfied in the framework of a unified approach. This testifies to potential effectiveness of applying chaos in communications and impels to persistently overcome the existing problems, namely: (i) necessity of constructing "precision" components of chaotic systems and "precision" systems as a whole, ensuring reproducibility of performance characteristics "from sample to sample"; (ii) high sensitivity of most proposed communication schemes to linear or nonlinear distortions in output circuits of transmitters and input circuits of receivers, and also directly in the communication channel; (iii) relatively low resistance to interference; (iv) absence of accomplished nodes and components; (v) strong difference of the proposed solutions from the mainstream of communication system development.

Some of these problems can be solved to a certain extent by means of choosing a modulation scheme [9], applying digital methods for forming, representing and processing chaotic signals [10], not using self-synchronization in the receiver [11], [12].

Here we consider an approach for solving the existing problems by means of using "direct chaotic" communication systems [13]–[15].

As is denoted in this paper, "direct chaotic" are the schemes where information is put into the chaotic signal generated directly in RF or microwave bands. Information signal is put, e.g., by means of modulating the transmitter parameters, or by means of modulating the chaotic signal by information after the chaotic signal is generated in the source. Consequently, the information signal is retrieved also in RF or microwave bands.

Note that direct chaotic schemes are potentially high-rate since chaotic signals are naturally wide-band.

Below we consider a practically realizable direct chaotic communication system, describe an experimental model providing high-rate data transmission, and discuss the results obtained with this model.

## II. DIRECT CHAOTIC SCHEME FOR INFORMATION TRANSMISSION

Analysis of potential direct chaotic communication systems prompts us to pay attention to schemes with the signals in the form of chaotic radio pulses. Chaotic radio pulse is a fragment of a direct chaotic signal whose duration exceeds the "quasi-period" of chaotic oscillations. Information transmitted by such pulse trains can be encoded by the location of the pulses in time domain, the pulse duration, the distance between the pulses, etc.

The sequence of chaotic radio pulses can be received either coherently, e.g., with a help of chaotic synchronization, or incoherently. Coherent receiver is approximately 6 dB more efficient in respect to signal-to-noise ratio. However, as will be shown below, incoherent reception in direct chaotic circuits of RF or microwave bands can be realized right now, while the coherent receiver is only a perspective.

In the basis of the proposed scheme for direct chaotic communications, three main ideas are laid which make it practically realizable: (1) chaotic source generates oscillations directly in the prescribed region of microwave

---

band; (2) information signal is put into the chaotic signal by means of forming a corresponding train of chaotic radio pulses; (3) information is retrieved with an incoherent receiver.

A block diagram of the communication scheme is shown in Fig. 1.

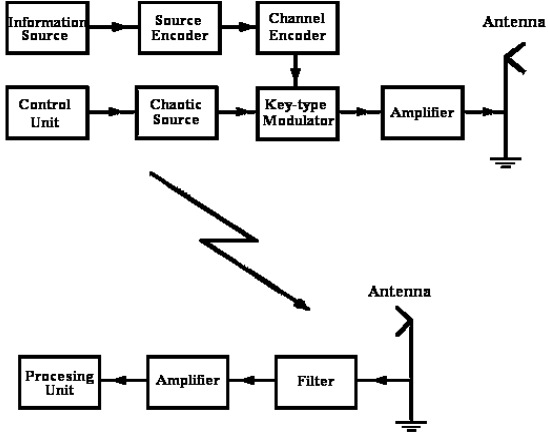

Fig. 1. Block diagram of direct chaotic communications.

The transmitter of the system is composed of a generator control unit, a chaotic source, that generates the signal directly in the communications frequency band, i.e., in RF or microwave band, a keying-type modulator, an amplifier, an antenna, an information source, a source encoder, and a channel encoder.

Chaotic source generates the signal with the center frequency $f_0$ and $\Delta f$ bandwidth. The center frequency and the bandwidth may be tuned by control unit. Modulator forms chaotic radio pulses and intervals between the pulses.

Information coming from information source is transformed by source encoder into a signal that is fed to the channel encoder, which in turn transforms it into a modulating signal that controls modulator. The duration of the formed chaotic pulses may be varied in the range $T \sim 1/\Delta f$ to $T \rightarrow \infty$.

The formed signal is passed through amplifier and is emitted in space with a wideband antenna. Information stream may be formed by means of varying the distance between the pulses, the pulse duration, the pulse rms amplitude, or a combination of these parameters.

For example, the stream nay be formed at a fixed pulse repetition rate and a fixed pulse duration. In this case, the presence of a pulse on a prescribed position in the stream is related to transmission of symbol "1", and the absence to symbol "0".

The receiver is composed of a wideband antenna, a filter passing the signal within the transmitter's frequency band, a low-noise amplifier, and a system of signal processing. The stream of chaotic radio pulses comes to antenna, is passed through filter and amplifier. Then the signal processing system recognizes the pulses, determines their parameters and their location in the stream, and retrieves useful information from the signal.

The processing system retrieves useful information from the signal by means of integration of the pulse power over the pulse time interval. Thus, incoherent receiver of the stream of chaotic radio pulses is realized in the system.

III. CHAOTIC RADIO PULSES

Chaotic radio pulse is a fragment of a direct chaotic signal with the duration exceeding the "quasi-period" of chaotic oscillations (Fig. 2).

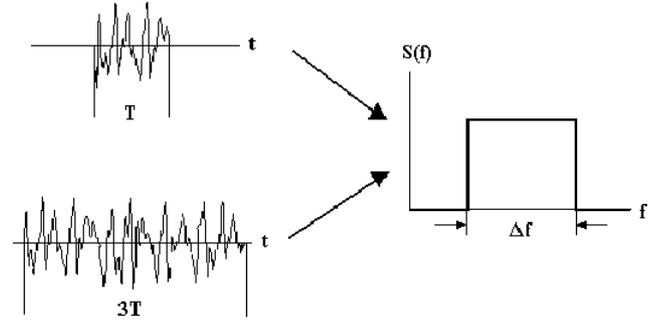

Fig. 2. Chaotic radio pulses (left) and their frequency band (right).

The bandwidth of chaotic radio pulse is determined by the frequency bandwidth of the original chaotic signal $\Delta f$ generated by the chaotic source, and is independent of the pulse duration in a wide range of pulse duration variation $\tau > 1/\Delta f$. So, Fig.2 illustrates that the chaotic pulses of different duration $T$ and $3T$ have the same form of power spectrum. This makes chaotic radio pulse essentially different from classical radio pulse filled with a fragment of a periodic carrier, whose bandwidth $\Delta f$ is determined by its period $T$:

$\Delta f = 1/T$

This property of conserving the pulse bandwidth with variations of its duration allows us to realize versatile schemes for pulse train modulation having fixed parameters of the transmitter output circuits and the receiver input circuits. For example, with variation of the pulse duration one needs not to change the bands of the input filter and the low-noise amplifier. While in the case of the classical radio pulse, a change of the pulse duration requires consequent changes of the frequency parameters of input and output circuits of the receiver and transmitter, respectively. This fact practically gives no chance to use different- duration pulses in the same communication system and impose essential restrictions on possible types of signal modulation, on variation of the transmission rate depending on the environment, etc.

Increasing the duration of chaotic radio pulse increases noise resistance of communications. In this case, the pulse energy increases, and this factor may be used to control the operation distance range without changing the transmitter's peak power.

To transmit an information bit in direct chaotic communication system, one might use single pulses as well as pulse sequences. In either case a stream of chaotic radio pulses is formed in time domain.

The duration of chaotic radio pulse and the mean duty cycle may be varied. This allows one to flexibly control the rate of information transmission by means of varying the pulse repetition rate and the mean power of the transmitted signal.

Due to above-mentioned features of chaotic radio pulse, these manipulations do not lead to substantial changes of spectral characteristics of the transmitted signal in respect to the signal of the chaotic source. No additional spectral components appear in the emitted signal.

## IV. MODULATION AND DEMODULATION

In the case of little duty cycle, the sequence of chaotic radio pulses interferes with the signals of traditional radio circuits only at very short time intervals. Thus, at the pulse repetition rate of $10^5$ pulse/sec and each pulse duration equal to $10^{-7}$ sec., duty cycle $S = 0.01$, i.e., the time of interference is 1% of the system operation time. In this case, the average emitted power is 100 times less than the average power on the intervals of chaotic radio pulse emission. If, for example, the transmitter power during emission is equal to 200 mW, then the average power is only 2 mW.

Periodic sequence of chaotic radio pulses carries no information due to regular character of pulse repetition.

In order to transmit information, additional signal manipulations are necessary that lead to modulation of the pulse train.

A number of different types of modulation can be used in direct chaotic communication systems. Let us describe two examples.

### A. The simplest system

In this case the pulse sequence is modulated as follows (Fig. 3). A sequence of chaotic radio pulses is formed from the original chaotic signal, with a constant repetition rate and constant pulse duration. At the output of the former each pulse is multiplied as a whole by +1, if symbol "1" is transmitted and by 0, if the symbol is "0".

### B. Pulse position modulation

In this case only the moment of emission of chaotic radio pulse is varied with respect to its nominal position. For example, in the system with 10 million pulses formed per second, the pulses are to be formed in every 100 nsec on the average. In the system with pulse position modulation, a pulse corresponding to symbol «0» is to be formed slightly earlier that the assigned 100 nsec position, and the pulse corresponding to symbol «1» slightly after such a position. Such type of modulation was used, for example, for ultrawide band system [16].

The above examples do not exhaust all possible modulation techniques.

## V. EXPERIMENTAL MODEL

An experimental model realizes the main functions of the simplest direct chaotic communication system shown in Fig. 1. The model layout is given in Fig. 4.

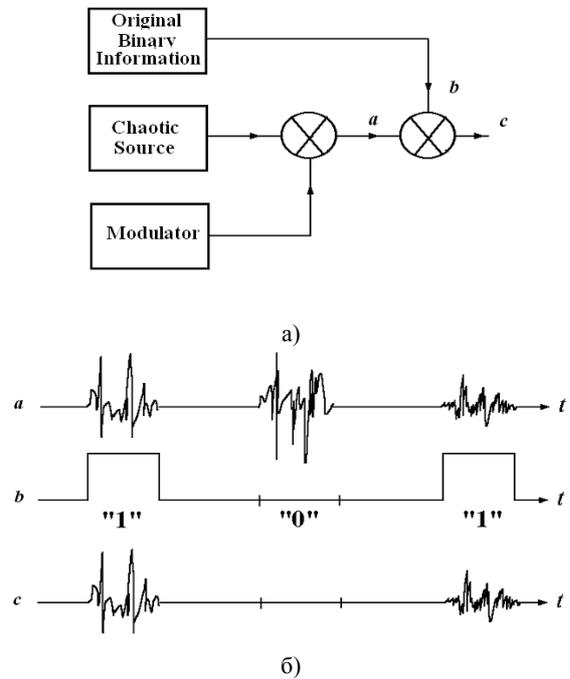

Fig. 3. Simplest modulation scheme. (a) and signal waveforms in various nodes of the circuit (b).

In the transmitter, chaotic generator produces continuous chaotic oscillations of ~ 1GHz wave band, that pass through modulator, that is controlled by pulse signal from pulse source, which combines the functions of information source, source encoder and channel encoder. In the modulator, a stream of chaotic radio pulses is formed from the chaotic signal, whose envelope corresponds to the control pulse stream. Then the chaotic radio pulses are amplified and emitted through antenna into space (communication channel). From the channel the signal comes to the receiver antenna, then it is passed through amplifier and is fed to the input of detector, which detects the incoming signal and filters out the RF component. Thus, the detector forms the envelope of radio pulses, a pulse signal corresponding to the controlling pulses of source. This pulse signal is depicted by oscilloscope.

Consider the elements of circuit in Fig. 4 in detail.

### A. Chaotic generator

The generator is developed on the basis of a three-point circuit with bipolar transistor as an active element. Classical single-transistor three-point circuits are designed to generate periodic signals. However, they can also generate chaotic oscillations in low-frequency RF and MW bands [17-19]. A feature of chaotic modes of such oscillators is *super wide bandwidth* of excited oscillations: the power spectrum extends both to the region of very low frequencies as well as to high frequencies, many times exceeding the main oscillation frequency. Use of such oscillators *as is* in direct chaotic communications is restrained due to distortions that super wideband signals undergo in real communication channels.

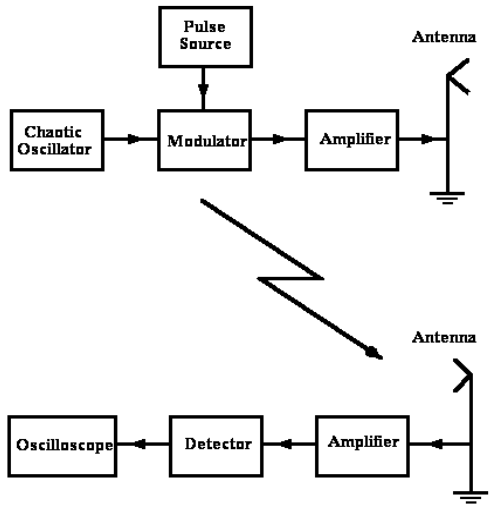

Fig. 4. Block diagram of the experimental model of direct chaotic communication system.

In order to obtain a band-limited chaotic signal, a resonant element is introduced in the oscillator feedback loop, which endows it the necessary frequency-selection properties, thus, creating conditions for generation preferably within the resonant element pass band. This principle of forming chaotic oscillations with prescribed frequency spectrum is proposed and investigated in [20-21].

In the experimental model of communication system this approach was used to develop a chaotic generator operating in 950–1050 MHz band [13,22]. The generator made in microstrip design on a 1 mm thick dielectric substrate with $\varepsilon = 2.8$ is shown in Fig. 5. The construction is based on capacitive three-point circuit. However, unlike this circuit a resonant element (RE) is introduced between collector (C) and emitter (E) in series with $C_1$, represented by a bounded strip-line resonator. The signal from the resonator output is fed to the outer circuit. The chaotic modes of the generator are tuned with variable capacitors $C_1$, $C_2$, and $C_3$ (4/20 pF) and by means of varying voltages $V_E$ and $V_C$. The pass band and non-uniformity of the resonator amplitude-frequency response determine, consequently, the band and non-uniformity of the power spectrum of the chaotic signal at the generator output.

Typical generator power spectrum is presented in Fig. 6. Output power is 5 mW.

### B. Modulator

The function of the modulator is to pass through or stop the chaotic signal from the generator judging from the voltage level at its control input. Pulse (two-level) signal is fed to the control input, and the voltage levels are adjusted so that one level corresponds to the signal pass mode, and the other to stop mode. The modulator is a micro-strip construction with *p-i-n* diodes. A specific feature of this construction is that the modulator opens and passes the chaotic signal when the lower-level signal is applied to the control input.

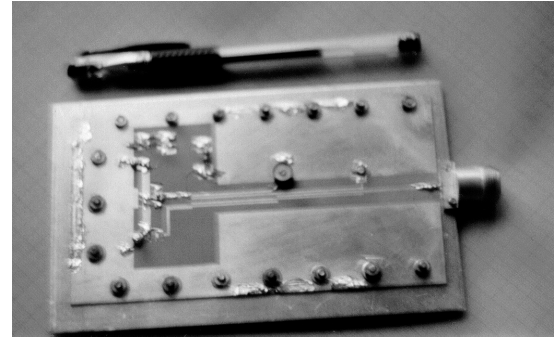

a)

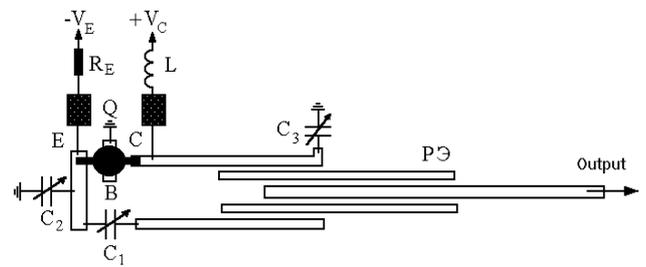

b)

Fig. 5. Chaotic generator for direct chaotic communications (a) and its topology (b).

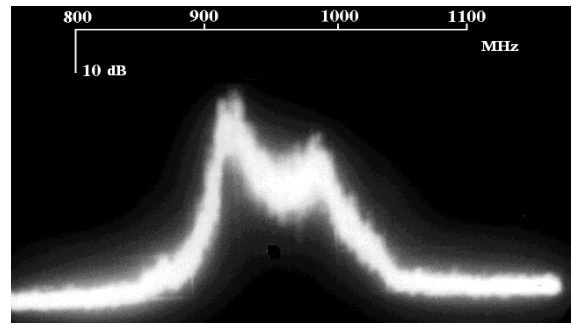

Fig. 6. Power spectrum of chaotic oscillations.

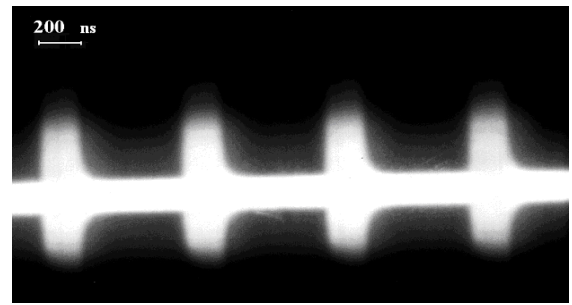

Fig. 7. Sequence of chaotic radio pulses.

## C. Pulse source

At first, in the experiments we used a pulse generator which allowed us to vary the pulse duration and repetition rate (duty cycle). Two test modes of control pulse streams were formed. The first mode with 150 ns pulse duration and 1/4 duty cycle provided the pulse rate of 1.67 million pulses per second (Mpps). The second mode with 50 ns pulse duration and 1/4 duty cycle provided the pulse rate of 5 Mpps.

Since this generator couldn't form pulses with duration less than 50 ns, in order to increase the rate of pulse transmission, we used a generator of 70 MHz sinusoidal oscillations as a pulse source. This signal was fed to the modulator, and chaotic radio pulses were formed with 7–10 ns duration and 1/1.4–1/2 duty cycle. The duration and duty cycle of the radio pulses were varied by means of varying the DC bias of the modulator. The rate of chaotic radio pulse repetition was fixed at 70 Mpulse/s.

## D. Amplifier

The amplifier has 30 dB gain for low signals in 950-1050 MHz range. In the experiments the average signal power at the amplifier output was 50–100 mW. The signal waveform at the amplifier output which is a sequence of chaotic radio pulses corresponding to the first test mode (150 ns) is presented in Fig. 7.

## E. Transmitting and receiving antennas

The transmitter signal was emitted to and was received from the communication channel by means of calibrated linear-polarization horn antennas. The width of the antenna directivity pattern at –3 dB level was about 60°. The closest side-lobe levels were less than –35 dB. Antennas orientations with respect to each other could be varied arbitrarily.

## F. Communication channel

The experiments were held in a laboratory room (280 m$^3$) with a large number of metal constructions (lab racks and stands). The distance between the transmitting and receiving antennas was ~ 10 m.

## G. Detector

The detector was intended to detect microwave signal, to filter out the chaotic carrier oscillations and to retrieve the envelope. The detector was made in the same box as the receiver amplifier, it was capable of retrieving the envelope at the chaotic pulse repetition rates of up to several tens of MHz. Oscilloscope was used to visually observe the microwave signal envelope retrieved by the detector.

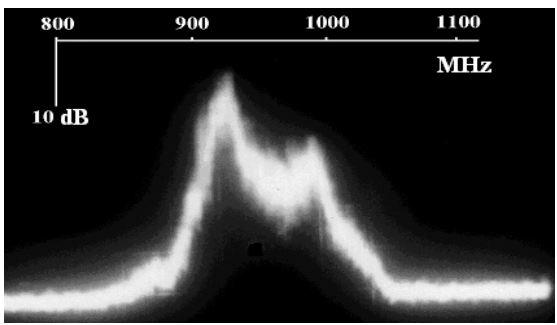

a)

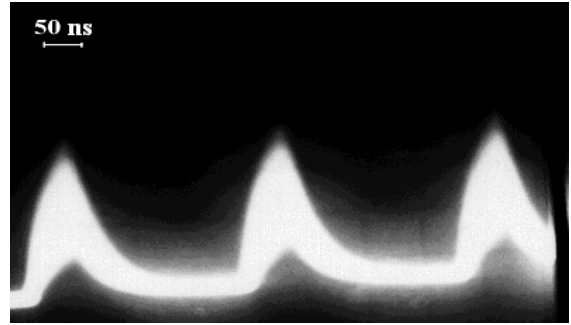

b)

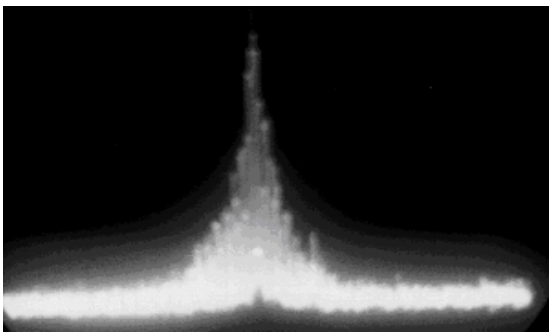

c)

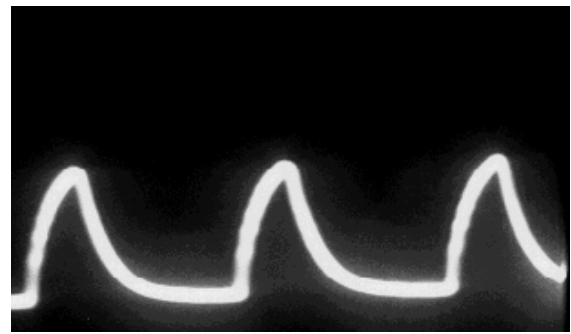

d)

Fig. 8. Power spectrums and waveforms of microwave signal envelopes in the receiver of the signals at the receiver input: (a)-(b) chaotic carrier; (c)-(d) the harmonic carrier.

## VI. EXPERIMENTAL RESULTS

The experiments included wireless transmission and receiving chaotic signals in different modulation modes, and investigation of the communication stability in respect to multipath propagation.

The first series of experiments was performed for the first test mode of control pulses (pulse duration 150 ns, duty cycle 1/4). The aim of the experiments was to test the experimental model elements and the model as a whole, and to verify the basic ideas.

The experiments were held with a typical operation mode of the chaotic generator, its power spectrum is given in Fig. 6. The stream of control pulses was fed to the modulator and was used to form the signal at the amplifier output, i.e., a sequence of chaotic radio pulses (Fig. 7).

The spectrum of the signal at the receiver input for the first test signal is shown in Fig. 8a. As follows from comparison of this spectrum with that in Fig. 6, the information component does not exhibit itself in the spectrum of the signal at the receiver input at the taken parameters of chaotic and modulating signals. The spectrum of the signal at the receiver input is close to that of the signal generated by the chaotic source.

The absence of signs of modulation in the spectrum of the signal coming from the channel is explained by the fact that the bandwidth of the carrying chaotic signal (100 MHz) is much broader than the frequency bandwidth determined by the chaotic radio pulse duration ($\Delta f \sim 1/T$, where $T$ is the radio pulse duration).

The wave form of the microwave signal envelope retrieved in the receiver testifies reliable reception of direct chaotic microwave signal and high quality of retrieving the form of the modulating signal. Note that the latter is not necessary, in general, for reliable transmission of digital data: for this purpose, detecting just the presence of pulse is sufficient.

Thus, in the first series of experiments principle functionality of the proposed direct chaotic circuit of communications in microwave band was demonstrated and 1.67 Mpps rate with 1/4 duty cycle was provided which corresponds to transmitting the sequence of zeros and ones with the rate of 3.34 Mpps with 1/2 duty cycle.

In the second series of experiments, a test mode with the stream of 50-ns control pulses (1/4 duty cycle) was used. The aim of the experiments was to increase the rate of information transmission in respect to the first series of experiments and to experimentally confirm conservation of the spectrum of the formed signal regardless of the presence or absence of modulation.

As in the first series, the experiments were performed for the typical mode of the chaotic generator (Fig. 6). The experiments demonstrated that as in the case of the previous series the information component did not exhibit itself in the power spectrum of the signal at the receiver input (Fig. 8a).

The wave form of the microwave signal envelope retrieved in the receiver is shown in Fig. 8b. Residual filling of pulses and intervals between the pulses by high-frequency chaotic component can be seen in the wave form. Nevertheless, the wave form in Figs. 8b indicates of reliable reception of direct chaotic microwave signal and of sufficiently high quality of retrieving the modulating signal form.

In order to compare the properties of information transmission using chaotic radio pulses and transmission using radio pulses with regular carrier, we performed additional experiments. In these experiments a harmonic mode of the generator was used for forming the stream of radio pulses. The frequency of oscillations was set at 950 MHz. The spectrum of the stream of radio pulses with harmonic carrier is presented in Fig. 8c. The second test mode of the control pulse stream was used for modulation. In this case, the signal at the modulator output has line spectrum and its effective width is determined by the pulse duration and for the discussed duration is ~ 10 MHz, which is much less than in the case of chaotic carrier. The form of the microwave signal envelope retrieved in the receiver (Fig. 8d) is in good agreement with the form of the modulating signal. Thus, in the second series of experiments we demonstrated direct chaotic information transmission in microwave band with the rate up to 10 Mbps and the constant power spectrum of chaotic signal.

In the third series of experiments the third test mode of control pulses was used with the pulse duration 7÷10 ns and duty cycle 1/1.4 ÷ 1/2. The aim of the experiments was to determine the limit of increase of the information transmission rate under condition of the spectrum independence of the presence or absence of modulation.

The spectrum of the stream of 10-ns duration and 1/1.4-duty cycle chaotic radio pulses was practically the same as the spectrum of unmodulated chaotic source signal and the spectrum of long-duration chaotic radio pulses (Fig. 8a). In the case of 7-ns duration and 1/2 duty cycle chaotic radio pulses a certain spectrum broadening and an increase of the spectrum envelope level outside the generation band were observed.

The aim of the fourth series of experiments was verification of the absence of fading in the case of multipath propagation.

The experiments were performed with the use of the first and second test modes of control pulse stream and a typical operation mode of the chaotic generator. We investigated the fading of the received signal in multi path environment. Measurements were made both in the presence and in the absence of direct beam between the transmitter and the receiver.

In the first case the transmitter and receiver antennas were directed at each other (Fig. 9a), which guaranteed the presence of direct beam. Additional signal propagation beams appeared due to reflections of electromagnetic waves from numerous metal constructions in the room.

In the second case the antennas were turned away so that the main lobe of the receiver antenna directivity pattern did not fit the main lobe of the transmitter antenna (Fig. 9b and 9c). In this case the receiver antenna received only the waves reflected by metal constructions in the room, and the total power level of the received signal could be 20–40 dB lower than in the first case.

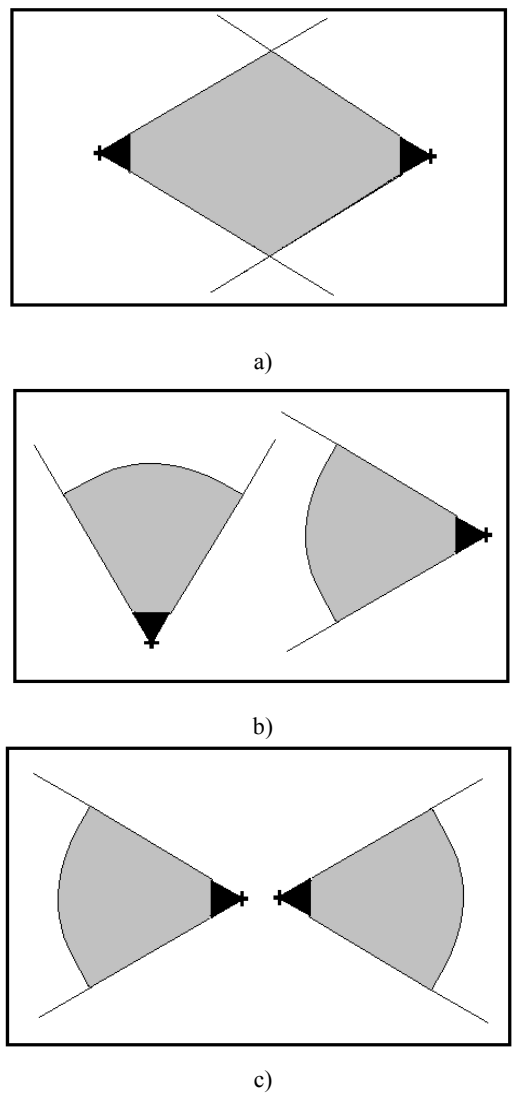

Fig. 9. Positions of the transmitter and receiver antennas in the experiments.

In all the measurements there was no noticeable interference fading, i.e., smooth turning the antenna axis didn't lead to sharp changes of the received signal intensity, which is characteristic of the situations when interference of the signals coming from different paths takes place. This was an expected effect associated with wide band character of the used chaotic signal.

Unexpected was that in the process of measurements we didn't observe any visible distortions of the form of the retrieved microwave signal envelope. Estimates indicate that this can take place only in the case when the typical path difference between different beams is below 3–5m.

Thus, in the experiments we didn't find negative effect of interference on the quality of reception in the model of direct chaotic communications.

## VII. CONCLUSIONS

Schemes for direct chaotic communications were proposed in the paper, based on three main ideas: (1) chaotic source generates chaotic oscillations directly in the prescribed region of microwave band; (2) information signal is put in the chaotic signal by means of forming the corresponding stream of chaotic radio pulses; (3) information is retrieved with a envelope detection.

The simplest scheme was realized in an experimental model of communications system, with which information was transmitted wirelessly in microwave band with the use of a novel carrier — dynamic chaos generated directly in the microwave band. Transmission rates equivalent to 3.34 Mbps, 10 Mbps and up to 70 Mbps were demonstrated.

Preliminary estimates as well as the results of the experiments showed that this communication scheme was sufficiently resistant to noise and other interference in wireless channel. The corresponding results will be published elsewhere.

The authors are grateful to Yu. V. Andreyev, L. V. Kuzmin and N. A. Maximov for useful discussion.

## REFERENCES

[1] L. M. Pecora and T. L. Caroll, "Synchronization in Chaotic Systems," *Phys. Rev. Lett.*, vol. 64, pp. 821-824, 1990.

[2] L. Kocarev, K. S. Halle, K. Eckert, L. Chua, and U. Parlitz, "Experimental Demonstration of Secure Communications via Chaotic Synchronization," *Int. J. Bifurcation Chaos*, vol. 2, pp, 709-713, 1992.

[3] U. Parlitz, L. Chua, L. Kocarev, K. Halle, and A. Shang, "Transmission of Digital Signals by Chaotic Synchronization," *Int. J. Bifurcation Chaos*, vol. 2, pp. 973-977, 1992.

[4] A. R. Volkovskii and N. F. Rulkov, "Synchronous Chaotic Response of Nonlinear Oscillating System as a Principle for the Detection of the Information Component of Chaos," *Sov. Tech. Phys. Lett.*, vol. 19., pp. 97-99, 1993.

[5] S. Hayes, C. Grebogi, and E. Ott, "Communicating with Chaos," *Phys. Rev. Lett.*, vol. 70, pp. 3031-3033, 1993.

[6] K. Cuomo and A. Oppenheim, "Circuit Implementation of Synchronized Chaos with Application to Communications," *Phys. Rev. Lett.*, vol. 71, pp. 65-68, 1993.

[7] Yu. L. Belskii and A. S. Dmitriev, "Information Transmission Using Chaos," *Journal of Communication Technology and Electronics*, vol. 38, pp. 1-5, 1993.

[8] H., Dedieu, M. P., Kennedy, and M. Hasler, "Chaos Shift Keying Modulation and Demodulation of a Chaotic Carrier Using Self-synchronizing Chua's Circuits," *IEEE Trans. Circuits and Systems – 1*, vol. 40, pp. 634-642, 1993.

[9] M. M. Sushchik, N. F. Rulkov, L. Larsen, L .S. Tsimring, H. D. I. Abarbanel, K. Yan, and A. R. Volkovskii, "Chaotic Pulse Position Modulation: a Robust Method of Communicating with Chaos," *IEEE Comm. Lett.*, vol. 4, pp. 128-130, 2000.

[10] Yu. V Andreyev, A. S. Dmitriev, L. V. Kuzmin, A. I. Panas, D. Yu. Puzikov, S. O. Starkov, S. V. Yemetz,

"Chaotic markers for communications", *Proc. NOLTA-2000*, Dresden, Germany, September 17 - 21, vol. 1, pp. 87-90, 2000.

[11]  U. Parlitz and S. Ergezinger, "Robust Communication Based on Chaotic Spreading Sequences," *Phys. Lett. A*, vol. 188, pp.146-150, 1994.

[12]  M. P. Kennedy and G. Columban, "Chaotic Modulation for Robust Digital Communications over Multipath Channels," *Int. J. Bifurcation Chaos*, vol. 10, pp. 695-719, 2000.

[13]  A. S. Dmitriev, B. Ye. Kyarginsky, N. A. Maximov, A. I. Panas, and S. O. Starkov, "Prospects of Direct Chaotic Communication Systems in RF and microwave bands," *Radiotechnika*, vol. 42, N 3, p. 9-20, 2000, (In Russian).

[14]  A. S. Dmitriev, B. Ye. Kyarginsky, N. A. Maximov, A. I. Panas, and S. O. Starkov, "Direct chaotic infromation transmission in microwave band,"*Preprint IRE RAS*, N 1 (625), 2000, (In Russian).

[15]  A. S. Dmitriev, B. Ye. Kyarginsky, A. I. Panas and S. O. Starkov, "Direct Chaotic Communications Schemes in Microwave Band," *Journal of Communication Technology and Electronics*, vol. 46, pp. 224-233, 2001.

[16]  R. Scholtz and M. Z. Win, "Impulse Radio:How it Work," IEEE Commun. Letters, Feb. 1998.

[17]  A. I. Panas, B. Ye. Kyarginsky, and N. A. Maximov, "Single-transistor microwave chaotic oscillator", *Proc. NOLTA-2000*, Dresden, Germany, vol. 2, pp. 445-448, September 17-21, 2000.

[18]  A. S. Dmitriev, V. P. Ivanov, and M. N., Lebedev, "Model of Transistor Oscillator with Chaotic Dynamic," *Soviet Journal of Communication Technology and Electronics*, vol. 33, N 10, pp. 169-172, 1988.

[19]  M. P. Kennedy, "Chaos in the Colpitts Oscillator," *IEEE Trans. on Circuits and System – 1*, vol. 41, N 11, pp. 771-775, 1994.

[20]  V. Burykin and A. Panas, "Chaotic synchronization of RF-generators," *Proc. NDES'97*, Moscow, Russia, pp. 548-553, 1997.

[21]  Yu. L. Belsky, A. S. Dmitriev, A. I. Panas, and S. O. Starkov, "Synthesis of band pass chaotic signals in self-oscillating systems," *Soviet Journal of Communication Technology and Electronics*, vol. 38, N 15, pp. 1-5, 1993.

[22]  A. S. Dmitriev, A. I. Panas, and S. O. Starkov, ". "Ring oscillating systems and their application to the systhesis of chaos generators," *Int. J. Bifurcation Chaos*. vol. 6, N 5, pp. 851-865, 1996.

**Alexander S. Dmitriev.** Leading Researcher of the Institute of Radioengineering and Electronics of RAS. Graduated the Moscow Institute of Physics & Technology (MIPT) (1971), Ph.D from MIPT (1974), Dr.Sci. degree from IRE RAS (1988), Professor of physics and mathematics (1995). Research interests: dynamic chaos and bifurcation phenomena; generation of dynamic chaos; information technologies based on dynamic chaos and nonlinear phenomena; applications of dynamic chaos in information networks and communication systems.

**Andrey I. Panas.** B. 1955. Graduated the Moscow Energy Institute in 1978, Department of Electron Devices. Since 1978 works with the Institute of Radioengineering and Electronics (IRE) of RAS, now in position of senior researcher. In 1988 received the Ph.D. degree from the IRE RAS for the thesis on chaotic oscillators based on two-mode oscillating systems. Research interests: dynamic chaos, chaotic synchronization, information processing and communication system.

**Sergey O. Starkov**. B. 1956, graduated the Moscow Institute for Physics and Technology (MIPT) in 1979. Received the Ph.D.degree from the same Institute in 1987. Senior researcher with the Institute of Radioengineering and Electronics of the Russian Academy of Sciences (IRE RAS). Research interests: generation of chaos and communications using chaotic carriers; chaotic synchronization; controlling chaos and encoding information using chaos; digital processing of chaotic signals.